\documentclass[10pt,twocolumn]{article}

\usepackage[letterpaper,top=0.85in,bottom=0.95in,left=0.7in,right=0.7in,
            columnsep=0.28in]{geometry}
\usepackage{amsmath,amssymb}
\usepackage{graphicx}
\usepackage{booktabs}
\usepackage{caption}
\usepackage{subcaption}
\usepackage{authblk}
\usepackage[table]{xcolor}  
\usepackage{adjustbox}    
\usepackage{microtype}
\usepackage[colorlinks=true,allcolors=blue!55!black]{hyperref}

\graphicspath{{./}{Plots/}}

\newcommand{\Norac}{N_{\mathrm{orac}}}

\title{\bfseries Maximum-Likelihood Amplitude Estimation for Quantum Monte Carlo
Integration on Trapped-Ion Hardware: Convergence, Noise-Floor Saturation,
Depth-Dependent Bias }

\author{Aditi Lal\textsuperscript{1}}
\author{Alex Khan\textsuperscript{1}\textsuperscript{2}}
\author{Rut Lineswala\textsuperscript{1}}
\author{Abhishek Chopra\textsuperscript{1}}
\affil{\textsuperscript{1}BosonQ Psi Corp, 235 Harrison St, Syracuse, NY 13202, USA}
\affil{\textsuperscript{2}National Quantum Laboratory (QLab), University of Maryland, College Park, MD 20742, USA}

\begin{document}

\twocolumn[
\begin{@twocolumnfalse}
\maketitle
\begin{abstract}
\noindent
Quantum Amplitude Estimation (QAE) promises to accelerate Monte Carlo integration
by replacing the classical $\varepsilon \propto N^{-1/2}$ sampling law with the
Heisenberg-limited $\varepsilon \propto N^{-1}$ scaling, but whether any part of
that advantage survives on present-day hardware is an empirical question. We report
a systematic experimental benchmark of a Maximum-Likelihood Amplitude Estimation
(MLAE) Quantum Monte Carlo pipeline on IonQ trapped-ion processors, comprising
$60$ hardware trials on IonQ Forte-1 across five configurations
($5$ and $9$ qubits, $100$--$200$ shots per depth, linear and exponential
amplification schedules, $\approx\!113$ device-hours), together with $100$ trials
on IonQ Aria-1 and Forte-1 noise models at $5$, $9$ and $19$ qubits. The pipeline combines controlled-$R_y$
rotational state preparation, Grover-based amplitude amplification and
maximum-likelihood inference, and is benchmarked against
$b_{\max}^{-1}\!\int_0^{b_{\max}}\!\sin^2\!x\,\mathrm{d}x$ with
$b_{\max}=\pi/5$.

Three results emerge. First, the pipeline is correct: in the noiseless limit it
attains $\varepsilon \propto N^{-0.88}$, close to the Heisenberg-scaling reference, reducing
the error $54$--$74\times$ over a $133\times$ increase in oracle calls. Second,
that scaling does \emph{not} survive on hardware. Every noisy backend saturates at
an error floor of $(2$--$5)\times10^{-2}$, with fitted exponents of
$0.04$--$0.15$ on Forte-1 linear schedules and $\alpha$ $\le\!0$ on the vendor noise models: beyond the optimal shallow depth, however, deeper amplification provides no sustained accuracy gain and instead returns to the noise floor.
 Forte-1 hardware also outperformed IonQ's own Aria-1 noise model
at every depth $m \ge 2$, indicating that vendor noise models are pessimistic
predictors of MLAE accuracy. The practical implication for uncertainty
quantification is that on current devices the useful operating point is a shallow, schedule-tuned amplification depth chosen so that the deepest
circuit lands near $p \approx 1/2$---not the deepest schedule the coherence budget
allows.
\end{abstract}
\vspace{1.2\baselineskip}
\end{@twocolumnfalse}
]

%======================================================================
\section{Introduction}

Monte Carlo integration underpins uncertainty quantification (UQ) across
computational mechanics, finance and reliability engineering
\cite{rebentrost2018,woerner2019,stamatopoulos2020}, and its cost is
governed by the central limit theorem: to reach accuracy $\varepsilon$ one needs
$N = \mathcal{O}(\varepsilon^{-2})$ samples. Quantum Amplitude Estimation
\cite{brassard2002} reduces this to $N = \mathcal{O}(\varepsilon^{-1})$ oracle
calls, and Montanaro \cite{montanaro2015} showed the speedup extends to a broad
class of Monte Carlo estimators. The canonical QAE construction relies on Quantum
Phase Estimation, whose ancilla register and long controlled-unitary chains make
it impractical on NISQ devices.

Maximum-Likelihood Amplitude Estimation (MLAE) \cite{suzuki2020} removes the phase
register entirely. It executes a family of Grover-amplified circuits at several
amplification depths $m_k$, records only the single-qubit measurement statistics,
and recovers the amplitude by maximising a combined likelihood. This trades an
expensive quantum register for classical post-processing and makes amplitude
estimation reachable on existing hardware; related ancilla-free variants include
iterative amplitude estimation \cite{grinko2021} and simplified approximate
counting \cite{aaronson2020}.

Whether MLAE actually delivers super-classical accuracy on today's devices is a
separate question from whether the algorithm is asymptotically correct. Two
effects work against it.
Tanaka et al.\ \cite{tanaka2021} analysed MLAE under depolarizing noise
and showed the estimator acquires a systematic bias; separately, Herbert
\cite{herbert2021} showed that the state-preparation step can erase the asymptotic
speedup altogether when the distribution is loaded by Grover--Rudolph
\cite{groverrudolph2002}.

Trapped-ion processors are a natural testbed for these circuits
\cite{wright2019}. They offer long
coherence times, high two-qubit gate fidelities, and---most relevant here---%
all-to-all connectivity, so the multi-controlled reflection at the heart of the
Grover operator incurs no SWAP-routing overhead.

This paper reports what a full MLAE Monte Carlo pipeline actually does on IonQ
hardware. Our contributions are:

\begin{enumerate}
\item A validated noiseless baseline establishing that the pipeline attains
      $\varepsilon \propto N^{-0.88}$ (Sec.~\ref{sec:noiseless}), so that any
      departure measured on hardware is attributable to the device and not to the
      implementation.
\item Measurement of the hardware error floor: all noisy backends saturate at
      $\approx 3\times10^{-2}$ with fitted exponents $\le 0.15$
      (Sec.~\ref{sec:saturation}).
\item Discovery and characterisation of a reproducible, strongly non-monotonic
      dependence of the saturated error on amplification depth, together with a
      quantitative noise model.
\end{enumerate}

%======================================================================
\section{Methodology}

\subsection{Benchmark problem and reference values}

We estimate the normalised integral
\begin{equation}
\begin{aligned}
  I \;&=\; \frac{1}{b_{\max}}\int_{0}^{b_{\max}} \sin^{2}\!x \,\mathrm{d}x \\
    \;&=\; \frac{1}{b_{\max}}\left(\frac{b_{\max}}{2}
            - \frac{\sin 2b_{\max}}{4}\right),
    \qquad b_{\max}=\frac{\pi}{5},
\end{aligned}
\end{equation}
whose analytic value is $I = 0.121586636$. The quantum circuit does not evaluate
$I$ but its midpoint-rule discretisation on $n_{\mathrm{div}} = 2^{n_b}$ nodes,
\begin{equation}
  I_{n_b} \;=\; \frac{1}{n_{\mathrm{div}}}\sum_{i=0}^{n_{\mathrm{div}}-1}
                \sin^{2}\!\left(\frac{b_{\max}}{n_{\mathrm{div}}}
                \left(i+\tfrac12\right)\right).
  \label{eq:disc}
\end{equation}
All errors in this paper are quoted against $I_{n_b}$, not $I$, so that
discretisation error is excluded and the reported quantity is purely the
estimation error of the quantum algorithm. The relevant reference values are
listed in Table~\ref{tab:truth}.

\begin{table}[t]
\centering
\small
\caption{Benchmark reference values. The register width is $n_b$ data qubits,
$n_b-2$ ancillas and one objective qubit, giving $2n_b-1$ total qubits. Errors
throughout are measured against $I_{n_b}$.}
\label{tab:truth}
\begin{adjustbox}{max width=\columnwidth}
\begin{tabular}{lrr}
\toprule
Total qubits  & $I_{n_b}$ & $|I_{n_b}-I|$ \\
\midrule
5    & $0.121197315$ & $3.89\times10^{-4}$ \\
9     & $0.121562320$ & $2.43\times10^{-5}$ \\
19  & $0.121586612$ & $2.38\times10^{-8}$ \\
\bottomrule
\end{tabular}
\end{adjustbox}
\end{table}

\subsection{State preparation and the Grover operator}

The register holds $n_b$ data qubits $|x\rangle$, one objective qubit $|m\rangle$
and $n_b-2$ ancillas for the multi-controlled reflection. Operator $\mathcal{P}$
places the data register in uniform superposition with a layer of Hadamards.
Operator $\mathcal{R}$ encodes the integrand through the objective qubit using one
$R_y$ and $n_b$ controlled-$R_y$ rotations with binary-weighted angles,
\begin{equation}
  \theta_i = 2\cdot 2^{i}\,\frac{b_{\max}}{n_{\mathrm{div}}},
  \qquad i = 0,\dots,n_b-1,
\end{equation}
so that
\begin{equation}
\begin{split}
  \mathcal{R}\,\mathcal{P}\,&|0\rangle^{\otimes n_b}|0\rangle \\
  &= \frac{1}{\sqrt{n_{\mathrm{div}}}}\sum_{i} |i\rangle
    \Big(\cos\phi_i|0\rangle + \sin\phi_i|1\rangle\Big), \\
  &\hphantom{{}={}}
    \phi_i = \frac{b_{\max}}{n_{\mathrm{div}}}\left(i+\tfrac12\right).
\end{split}
\end{equation}
Measuring the objective qubit therefore yields $|1\rangle$ with probability
$a = \sin^2\theta_a = I_{n_b}$.

The Grover operator is $\mathcal{Q} = \mathcal{R}\,\mathcal{P}\,\mathcal{S}_0\,
\mathcal{P}^{\dagger}\mathcal{R}^{\dagger}\mathcal{S}_{\chi}$, with
$\mathcal{S}_{\chi}$ a $Z$ on the objective qubit and $\mathcal{S}_0$ the
reflection about $|0\rangle$ implemented as an $n_b$-controlled NOT through the
ancilla chain. Applying $\mathcal{Q}$ $m$ times gives
\begin{equation}
  p_m \;=\; \Pr[\,1\,|\,m\,] \;=\; \sin^{2}\!\big((2m+1)\theta_a\big).
  \label{eq:signal}
\end{equation}

\subsection{Maximum-likelihood inference}

For a schedule $\{m_k\}_{k=0}^{M}$ with $N_k$ shots at depth $m_k$ and $h_k$
observed $|1\rangle$ outcomes, the combined log-likelihood is
\begin{equation}
\begin{split}
  \ln L(\theta) &= \sum_{k=0}^{M}
   \Big[\, 2h_k \ln\big|\sin\big((2m_k+1)\theta\big)\big| \\
  &\hphantom{{}={}}
   + 2(N_k-h_k)\ln\big|\cos\big((2m_k+1)\theta\big)\big| \Big],
\end{split}
  \label{eq:ll}
\end{equation}
maximised over $\theta$ by grid search. After each step the search interval is
narrowed to $\hat{p}_k \pm 5\,\sigma_{\mathrm{CR}}$, where $\sigma_{\mathrm{CR}}$
is the Cram\'er--Rao bound from the Fisher information
\begin{equation}
\begin{split}
  \mathcal{I}(p) &= \sum_{k=0}^{M}\frac{N_k(2m_k+1)^{2}}{p(1-p)}, \\
  \sigma_{\mathrm{CR}} &= \mathcal{I}(p)^{-1/2}.
\end{split}
  \label{eq:crb}
\end{equation}
The reported estimate after step $k$ uses all depths $0,\dots,k$; the Fisher
weight of depth $k$ grows as $(2m_k+1)^2$, so the newest and deepest circuit
dominates the fit.

\subsection{Amplification schedules and oracle accounting}

Two schedules were run. The \emph{linear} schedule (LIS) uses $m_k = k$ for
$k=0,\dots,6$; the \emph{exponential} schedule (EIS) uses
$m_k \in \{0,1,2,4,8,16,32\}$. The cumulative oracle cost through step $M$ is
\begin{equation}
  \Norac(M) = \sum_{k=0}^{M} N_k\,(2m_k+1),
  \label{eq:orac}
\end{equation}
giving $\Norac = \{100,400,900,1600,2500,3600,4900\}$ for LIS and
$\{100,400,900,1800,3500,6800,13300\}$ for EIS at $100$ shots per depth. The
Heisenberg-limited reference used throughout is $\varepsilon_{\mathrm{th}} =
1/\Norac$.

\subsection{Platforms and experimental results}

Table~\ref{tab:config} lists every run. Hardware executions used IonQ Forte-1;
noise-model executions used IonQ's Aria-1 and Forte-1 device models.

\begin{table*}[t]
\centering
\small
\caption{Experimental campaign. Wall-clock times are as recorded for each
campaign. Hardware runs total $6754$ minutes $\approx 113$ device-hours.}
\label{tab:config}
\begin{adjustbox}{max width=\textwidth}
\begin{tabular}{llrrrrl}
\toprule
Backend & Type & Qubits & Schedule & Shots & Trials & Wall clock \\
\midrule
Noiseless simulator & statevector & 5  & EIS & 100 & $2\times20$ & --- \\
IonQ Aria-1 model   & noise model & 5  & EIS & 100 & 30 & 208\,min / 100 tr. \\
IonQ Forte-1 model  & noise model & 5  & EIS & 100 & 30 & 673.5\,min / 40 tr. \\
IonQ Aria-1 model   & noise model & 9  & LIS & 100 & 10 & 13\,min 34\,s \\
IonQ Aria-1 model   & noise model & 19 & LIS & 100 & 30 & 197.7\,min / 50 tr. \\
\midrule
\textbf{IonQ Forte-1} & \textbf{hardware} & 5 & LIS & 100 & 10 & 1524\,min 59\,s \\
\textbf{IonQ Forte-1} & \textbf{hardware} & 9 & LIS & 100 & 10 & 830\,min 18\,s \\
\textbf{IonQ Forte-1} & \textbf{hardware} & 9 & EIS & 100 & 10 & 645\,min 30\,s \\
\textbf{IonQ Forte-1} & \textbf{hardware} & 9 & LIS & 150 & 10 & 706\,min 19\,s \\
\textbf{IonQ Forte-1} & \textbf{hardware} & 9 & LIS$^{\dagger}$ & 200 & 20 & 3047\,min 23\,s \\
\bottomrule
\multicolumn{7}{l}{\footnotesize $^{\dagger}$ truncated at $m=5$. \quad
$^{\ddagger}$ truncated at $m=16$.}
\end{tabular}
\end{adjustbox}
\end{table*}

\subsection{Error metric}

For each trial and each schedule step $k$ we record
\begin{equation}
  \varepsilon_k = \big|\,\hat{p}_k - I_{n_b}\,\big|,
\end{equation}
the absolute deviation of the maximum-likelihood amplitude estimate from the
discretised reference of Eq.~\eqref{eq:disc}. All tabulated values are means over
trials; quoted uncertainties are sample standard deviations unless stated as
standard errors.

%======================================================================
\section{Circuit complexity on trapped-ion hardware}
\label{sec:cost}

\begin{figure*}[t]
\centering
\includegraphics[width=\textwidth]{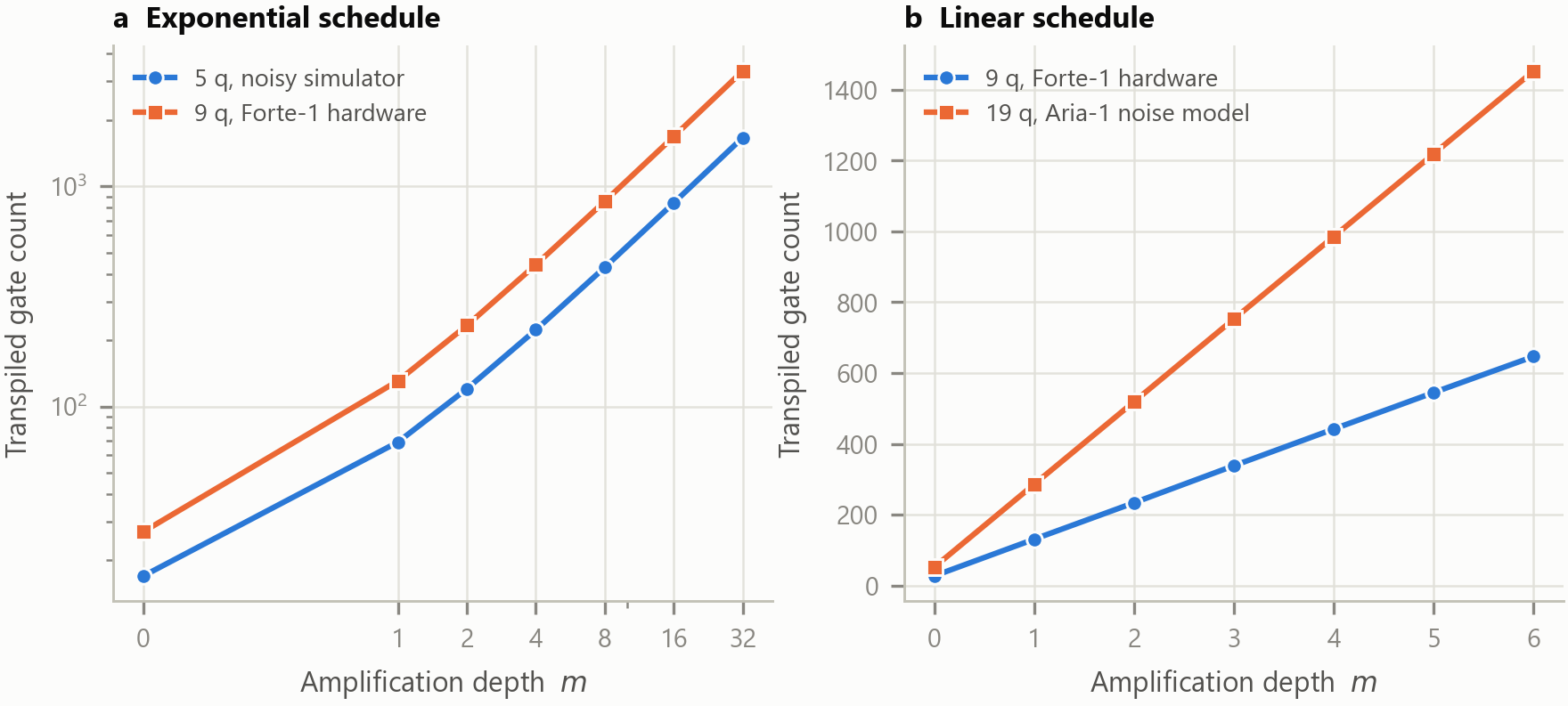}
\caption{Transpiled circuit cost. (a) Exponential schedule, logarithmic axes.
(b) Linear schedule: gate count is strictly linear in $m$ for both register
widths, with no routing overhead}
\label{fig:cost}
\end{figure*}

Table~\ref{tab:gates} and Fig.~\ref{fig:cost} give the transpiled circuit cost.
Two features matter for what follows.

First, the cost is exactly linear in the number of Grover iterations, as it must
be: on the $9$-qubit register %a {\color{green}{least-squares fit of total gates against $m$ gives $103.4\,m + 27.3$ ($R^2 = 0.99999$), and on $19$ qubits $233.5\,m + 52.2$ ($R^2 > 0.99999$).}}
there is
no super-linear blow-up from routing---a direct consequence of all-to-all
connectivity, which lets the $n_b$-controlled reflection compile without SWAP
insertion.

Second, the depth reached is substantial. The deepest circuit executed on hardware
($9$ qubits, EIS, $m=32$) carries $3337$ gates; the deepest $5$-qubit circuit in
the campaign carries $1664$ gates at circuit depth $1175$. These are the circuits whose
measurement statistics Sec.~\ref{sec:saturation} shows to be information-free.

\begin{table*}[t]
\centering
\small
\caption{Mean transpiled gate counts. Depth is reported for the $5$-qubit
register, where it was recorded per trial.}
\label{tab:gates}
\begin{adjustbox}{max width=\textwidth}
\begin{tabular}{lrrrrrrr}
\toprule
& \multicolumn{7}{c}{Grover iterations $m$} \\
\cmidrule(l){2-8}
Configuration & 0 & 1 & 2 & 4 & 8 & 16 & 32 \\
\midrule
5 q, gates (noisy sim.)   & 17 & 68.9  & 120.3 & 222.8 & 428.6 & 840.8  & 1663.7 \\
5 q, depth (noisy sim.)   & 14 & 50    & 87    & 160   & 304   & 596    & 1175 \\
9 q, gates (hardware, EIS)& 27 & 130.7 & 233.4 & 440.4 & 854.8 & 1682.2 & 3337.2 \\
\midrule
& 0 & 1 & 2 & 3 & 4 & 5 & 6 \\
\midrule
9 q, gates (hardware, LIS) & 27 & 130.6 & 233.6 & 338.3 & 441.9 & 544.9  & 646.7 \\
19 q, gates (noise model)  & 52 & 285.9 & 519.1 & 753.0 & 985.8 & 1219.3 & 1453.1 \\
\bottomrule
\end{tabular}
\end{adjustbox}
\end{table*}

%======================================================================
\section{Results}

\subsection{Noiseless baseline: the pipeline attains near-Heisenberg scaling}
\label{sec:noiseless}

Before attributing anything to hardware we verify the implementation. Two
independent $20$-trial noiseless statevector runs on the $5$-qubit register
(Table~\ref{tab:main}) reduce the error from
$2.2\times10^{-2}$ at $\Norac = 100$ to $4.1\times10^{-4}$ at
$\Norac = 13\,300$. A power-law fit
$\varepsilon \propto \Norac^{-\alpha}$ gives
$\alpha = 0.841$ and $\alpha = 0.923$, bracketing the classical $\alpha = 0.5$ and
approaching the Heisenberg $\alpha = 1$.

We further verified the estimator end-to-end: taking the raw per-shot hit counts
recorded for one $20$-trial run and re-running the likelihood of Eq.~\eqref{eq:ll} from scratch reproduced the tabulated error sequence
$\{0.022102, 0.007062, 0.004312, 0.003046, 0.001546, \\ 0.000468, 0.000406\}$ to
within $7\times10^{-6}$ at every depth---some $500\times$ tighter than the
trial-to-trial standard error of $3.7\times10^{-3}$.

\subsection{Every noisy backend saturates at an error floor}
\label{sec:saturation}

Table~\ref{tab:alpha} give the central negative result.
On every noisy backend, the error eventually returns to and remains near a floor in the range $(2$--$5)\times10^{-2}$, , although a pronounced shallow-depth minimum occurs at \(m=3\) in the hardware configurations.

The contrast with the theoretical bound is stark. Over the EIS schedule the
Heisenberg reference $1/\Norac$ falls by a factor of $133$; the measured error on
Forte-1 hardware falls by a factor of $1.2$--$1.3$ on the linear schedule and not
at all on the exponential one. Fitted exponents are $\alpha = 0.04$--$0.15$ on
hardware and $\alpha \le 0$ on all four vendor noise-model configurations, against
$\alpha = 0.88$ noiseless. Expressed as a ratio to the Heisenberg reference
$1/\Norac$, the measured error on the $9$-qubit hardware run degrades from
$\approx 3\times$ at $\Norac = 100$ to $440\times$ at $\Norac = 13\,300$.

The practical reading is that For this benchmark on Forte-1, amplification depths beyond \(m\approx3-4\) provided no sustained accuracy improvement and generally increased the estimation error.. The
additional $12\,400$ oracle calls spent going from $\Norac = 900$ to
$\Norac = 13\,300$ on the $9$-qubit exponential schedule did not merely fail to
improve the estimate: the error rose, from $2.2\times10^{-2}$ to
$3.3\times10^{-2}$, within a run consuming $645$ minutes of device time.

\begin{figure*}[t]
    \centering
    \begin{subfigure}{0.48\linewidth}
        \centering
        \includegraphics[width=\linewidth]{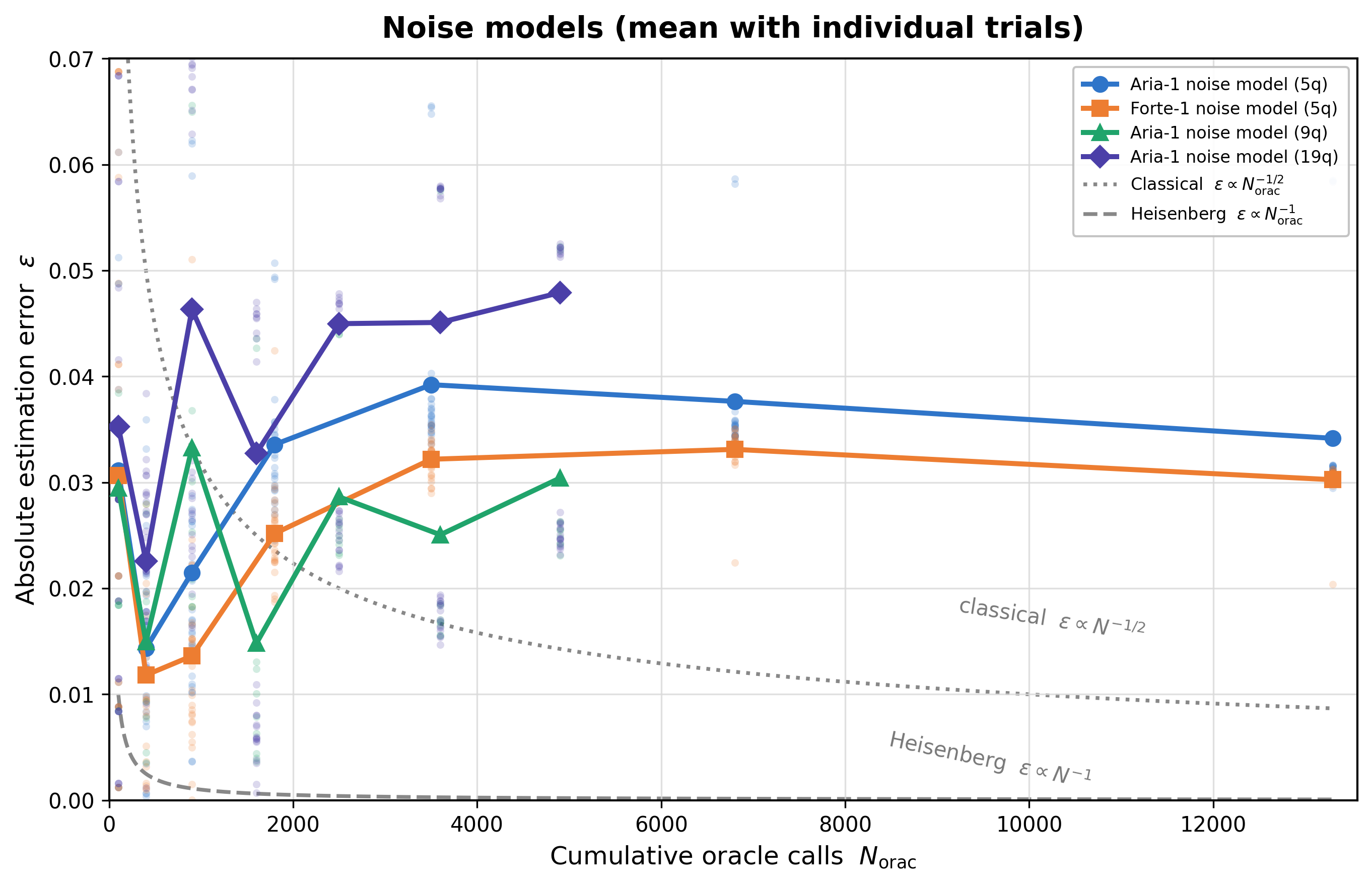}
    \end{subfigure}
    \hfill
    \begin{subfigure}{0.48\linewidth}
        \centering
        \includegraphics[width=\linewidth]{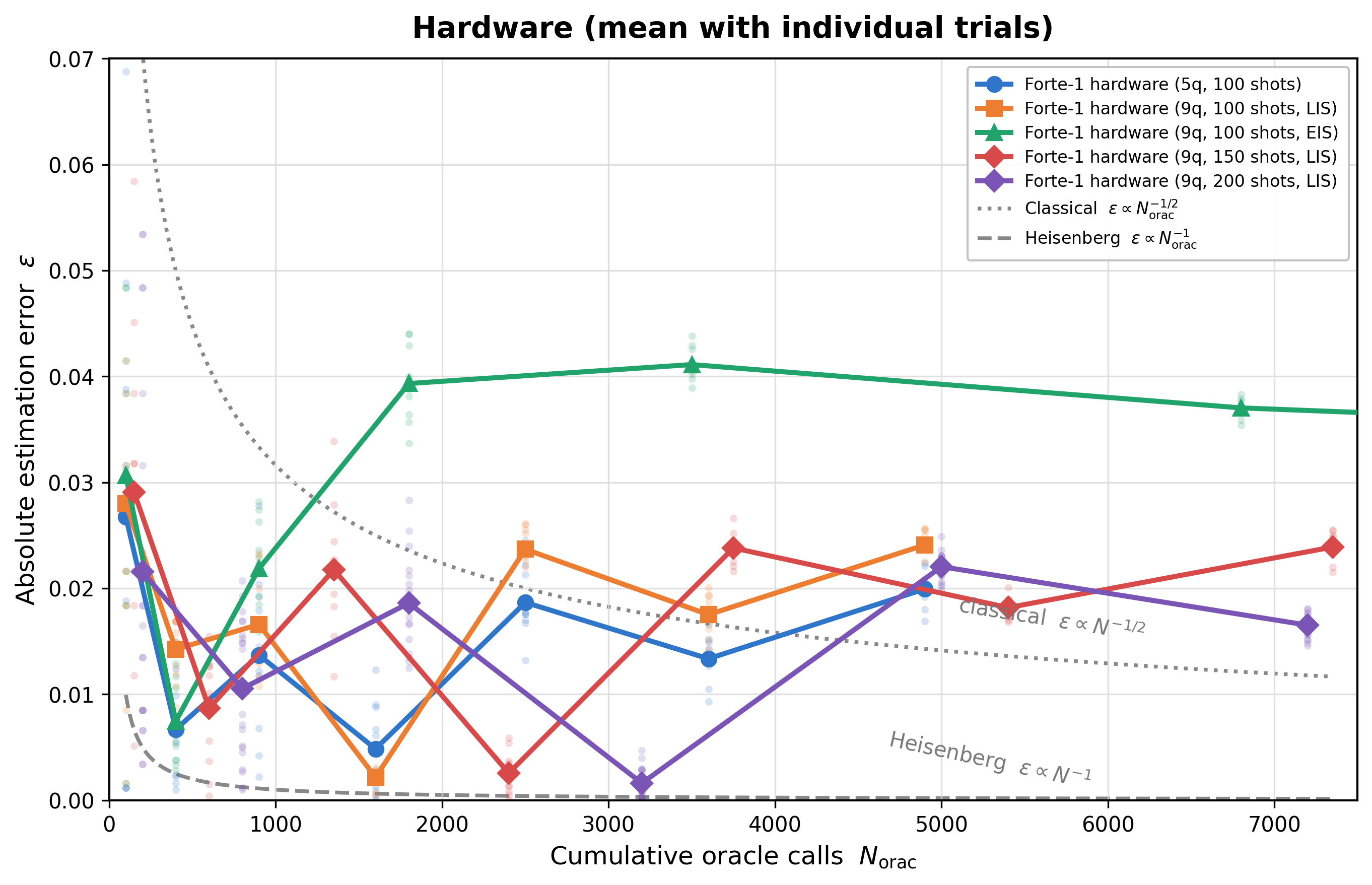}
    \end{subfigure}

    \caption{Estimation error as a function of cumulative oracle calls for the IonQ noise-model and hardware experiments.}
    \label{fig:ionq_results}
\end{figure*}

\begin{table*}[t]
\centering
\footnotesize
\caption{Mean absolute error $\varepsilon$ by schedule step. Upper block:
exponential schedule. Lower block: linear schedule. Bold rows are IonQ Forte-1
hardware.}
\label{tab:main}
\begin{adjustbox}{max width=\textwidth}
\begin{tabular}{lrrrrrrr}
\toprule
\multicolumn{8}{c}{\textit{Exponential schedule} \quad $m = 0,1,2,4,8,16,32$} \\
\midrule
$\Norac$ (100 shots) & 100 & 400 & 900 & 1800 & 3500 & 6800 & 13300 \\
$1/\Norac$ & $10^{-2}$ & $2.5{\cdot}10^{-3}$ & $1.1{\cdot}10^{-3}$ & $5.6{\cdot}10^{-4}$ & $2.9{\cdot}10^{-4}$ & $1.5{\cdot}10^{-4}$ & $7.5{\cdot}10^{-5}$ \\
\midrule
Noiseless sim.\ (run 1) & 0.02210 & 0.00707 & 0.00431 & 0.00304 & 0.00155 & 0.00047 & 0.00041 \\
Noiseless sim.\ (run 2) & 0.02512 & 0.00786 & 0.00548 & 0.00443 & 0.00114 & 0.00041 & 0.00034 \\
Aria-1 model, 5 q       & 0.03110 & 0.01434 & 0.02150 & 0.03357 & 0.03922 & 0.03765 & 0.03416 \\
Forte-1 model, 5 q      & 0.03066 & 0.01181 & 0.01362 & 0.02518 & 0.03218 & 0.03312 & 0.03025 \\
\textbf{Forte-1 hw, 9 q}& \textbf{0.03067} & \textbf{0.00745} & \textbf{0.02186} & \textbf{0.03933} & \textbf{0.04112} & \textbf{0.03703} & \textbf{0.03306} \\
\midrule[\heavyrulewidth]
\multicolumn{8}{c}{\textit{Linear schedule} \quad $m = 0,1,2,3,4,5,6$} \\
\midrule
$\Norac$ (100 shots) & 100 & 400 & 900 & 1600 & 2500 & 3600 & 4900 \\
$1/\Norac$ & $10^{-2}$ & $2.5{\cdot}10^{-3}$ & $1.1{\cdot}10^{-3}$ & $6.3{\cdot}10^{-4}$ & $4.0{\cdot}10^{-4}$ & $2.8{\cdot}10^{-4}$ & $2.0{\cdot}10^{-4}$ \\
\midrule
Aria-1 model, 9 q  & 0.02944 & 0.01494 & 0.03325 & 0.01480 & 0.02863 & 0.02506 & 0.03041 \\
Aria-1 model, 19 q & 0.03524 & 0.02257 & 0.04633 & 0.03274 & 0.04498 & 0.04510 & 0.04791 \\
\textbf{Forte-1 hw, 5 q, 100 sh} & \textbf{0.02676} & \textbf{0.00669} & \textbf{0.01370} & \textbf{0.00485} & \textbf{0.01867} & \textbf{0.01336} & \textbf{0.01994} \\
\textbf{Forte-1 hw, 9 q, 100 sh} & \textbf{0.02800} & \textbf{0.01425} & \textbf{0.01660} & \textbf{0.00217} & \textbf{0.02369} & \textbf{0.01752} & \textbf{0.02410} \\
\textbf{Forte-1 hw, 9 q, 150 sh} & \textbf{0.02906} & \textbf{0.00868} & \textbf{0.02175} & \textbf{0.00257} & \textbf{0.02381} & \textbf{0.01818} & \textbf{0.02390} \\
\textbf{Forte-1 hw, 9 q, 200 sh} & \textbf{0.02157} & \textbf{0.01054} & \textbf{0.01861} & \textbf{0.00161} & \textbf{0.02204} & \textbf{0.01653} & --- \\
\bottomrule
\end{tabular}
\end{adjustbox}
\end{table*}

\begin{table}[t]
\centering
\footnotesize
\caption{Power-law exponents $\alpha$ from $\varepsilon \propto \Norac^{-\alpha}$.
Classical sampling gives $\alpha = 0.5$; the Heisenberg limit gives
$\alpha = 1.0$.}
\label{tab:alpha}
\begin{adjustbox}{max width=\columnwidth}
\begin{tabular}{llrr}
\toprule
Configuration & Schedule & $\alpha$ & $\varepsilon_{\mathrm{first}}/\varepsilon_{\mathrm{last}}$ \\
\midrule
Noiseless simulator, 5 q (run 1) & EIS & $0.841$ & $54.3\times$ \\
Noiseless simulator, 5 q (run 2) & EIS & $0.923$ & $73.7\times$ \\
\midrule
Aria-1 noise model, 5 q   & EIS & $-0.113$ & $0.9\times$ \\
Forte-1 noise model, 5 q  & EIS & $-0.106$ & $1.0\times$ \\
Aria-1 noise model, 9 q   & LIS & $-0.026$ & $1.0\times$ \\
Aria-1 noise model, 19 q  & LIS & $-0.115$ & $0.7\times$ \\
\midrule
Forte-1 hardware, 5 q, 100 shots & LIS & $0.042$ & $1.3\times$ \\
Forte-1 hardware, 9 q, 100 shots & LIS & $0.081$ & $1.2\times$ \\
Forte-1 hardware, 9 q, 150 shots & LIS & $0.041$ & $1.2\times$ \\
Forte-1 hardware, 9 q, 200 shots & LIS & $0.145$ & $1.3\times$ \\
Forte-1 hardware, 9 q, 100 shots & EIS & $-0.164$ & $0.9\times$ \\
\bottomrule
\end{tabular}
\end{adjustbox}
\end{table}

\subsection{Hardware outperforms the vendor noise models}
\label{sec:models}

At matched register width, schedule and shot count ($9$ qubits, LIS, $100$ shots),
Forte-1 hardware produced a lower error than IonQ's Aria-1 noise model at every
depth $m \ge 2$, by factors of $2.0$ at $m=2$, $6.8$ at $m=3$, $1.2$ at $m=4$,
$1.4$ at $m=5$ and $1.3$ at $m=6$ . The two are not the
same device model---Forte-1 hardware against an Aria-1 noise model---so this is
not a like-for-like validation of a noise model against its own device. It does,
however, establish that publicly available IonQ noise models are a pessimistic
proxy for MLAE accuracy on Forte-1, and that pre-hardware simulation studies based
on them will understate achievable performance.

Register width degrades accuracy independently of depth. Under the same Aria-1
model and schedule, moving from $9$ to $19$ qubits raised the mean error at every
depth, by $20\%$ to $121\%$, consistent with the
$2.2\times$ larger gate count reported in Table~\ref{tab:gates}.

%======================================================================
\newpage
\section{Discussion}

\subsection{What the noise floor means for uncertainty quantification}

A UQ practitioner choosing between classical and quantum Monte Carlo cares about
one number: the accuracy achievable per unit of sampling effort. On the evidence
here, MLAE on Forte-1 delivers $\varepsilon \approx 1.6\times10^{-3}$ at
$\Norac = 3200$ in the best case, and $\approx 3\times10^{-2}$ generically. A
classical Monte Carlo estimator of the same integral reaches $\varepsilon =
1.6\times10^{-3}$ with $\sigma^2/\varepsilon^2 \approx 4.1\times10^{4}$ samples,
where $\sigma^2 = I(1-I) = 0.107$. Measured in queries, then, the quantum run is
about $13\times$ cheaper---but only at that single tuned operating point, and
nowhere else in the schedule.

Two caveats keep this from being a speedup. First, a quantum oracle call and a
classical sample are not comparable units of work: the $200$-shot campaign that
produced the $1.6\times10^{-3}$ result occupied $3047$ minutes of device time,
against microseconds for $4\times10^{4}$ classical samples. Query-complexity
parity at one point on the schedule is a statement about asymptotics, not about
time-to-solution today. Second, the saturation is a \emph{bias} floor, not a
variance floor, so it is not removed by taking more shots: the $200$-shot
campaign saturates at the same level as the $100$-shot campaign
(Table~\ref{tab:main}).

\section{Limitations}
\label{sec:limits}

\begin{enumerate}
\item \textbf{One integrand, one amplitude.} All results are for a single smooth
      benchmark at a single $b_{\max}$, hence a single $\theta_a$. The
      depth-dependent structure is a function of
      $\theta_a$; its \emph{existence} should generalise, but the specific
      location of the minimum at $m = 3$ will not.
\item \textbf{Trial counts.} Hardware campaigns comprise $10$--$20$ trials per
      configuration, the observed depth structure was reproducible across the four hardware configurations.
\item \textbf{Not a like-for-like noise-model validation.} Sec.~\ref{sec:models}
      compares Forte-1 hardware against an Aria-1 noise model, the only matched
      register/schedule/shot pairing available.
\item \textbf{Device drift.} Hardware runs span many hours to several days of
      wall clock (Table~\ref{tab:config}); no interleaved calibration reference
      was recorded, so slow drift is folded into the reported scatter.
\item \textbf{State preparation.} The rotational encoding used here exploits an
      analytically integrable integrand. It is not a general distribution loader,
      and no claim about asymptotic advantage for arbitrary distributions follows
      from these results \cite{herbert2021}.
\item \textbf{Error mitigation-IonQ debiasing:} IonQ provides a compiler-level error-mitigation technique called debiasing, which uses symmetry-based circuit variants to reduce hardware-induced errors. For direct IonQ Cloud jobs, debiasing is enabled by default only for runs with $\geq$ 500 shots. Our experiments used 100--200 shots per depth; therefore, IonQ debiasing was not applied to the reported hardware results.
\end{enumerate}

%======================================================================
\section{Conclusions}

We benchmarked a maximum-likelihood amplitude estimation Quantum Monte Carlo
pipeline on IonQ trapped-ion hardware across five configurations and roughly
$113$ device-hours.

The pipeline is correct: noiselessly it achieves $\varepsilon \propto N^{-0.88}$,
close to the Heisenberg limit. That advantage does not survive contact with
hardware. Every noisy backend saturates in the range $(2$--$5)\times10^{-2}$ with
fitted exponents no better than $0.15$, so the $133\times$ increase in oracle calls
across the exponential schedule bought a factor of $1.2$ in accuracy. On current
devices, MLAE is not delivering a quantum sampling advantage for this problem.

Two findings are more encouraging. Forte-1 outperformed IonQ's own Aria-1 noise
model at every depth $m \ge 2$, so simulator-based projections understate what the
hardware does. The saturated error is strongly structured in amplification depth, with a reproducible minimum at \(m=3\) across all four hardware configurations; for the 9-qubit configurations, this minimum is roughly an order of magnitude below the subsequent error floor We showed that a single-parameter depolarizing model reproduces this structure while a
shot-noise-only model does not, and traced it to the dependence of the injected
bias on $|p_m^{\mathrm{ideal}} - 1/2|$.

The route forward suggested by these data is not deeper circuits. It is
adaptive, shallow schedules that place each amplification depth where the
Grover signal is least sensitive to depolarization---worth an order of magnitude
here at zero additional oracle cost---combined with error mitigation targeted at
the bias, not the variance, of the likelihood.

%======================================================================
\section{Acknowledgements}
Access to IonQ Forte-1 was provided by the
National Quantum Laboratory (QLab) at the University
of Maryland.
%======================================================================

%======================================================================
{\small
\raggedright

}

\end{document}